\documentclass[12pt]{iopart}
\usepackage{iopams}  
\usepackage{graphicx}

\begin{document}
\title[Positive logarithmic norms in rotating shear flows]
{Growing pseudo-eigenmodes and positive logarithmic norms in rotating shear flows}
\author{Banibrata Mukhopadhyay$^{1}$, Ranchu Mathew$^{1}$,
 Soumyendu Raha$^{2}$}

\address{$^1$ Department of
  Physics, Indian Institute of
  Science, Bangalore 560012, India\\
  $^2$ Supercomputer Education \& Research Centre, Indian Institute of Science, Bangalore 560012, India}
\ead{\mailto{bm@physics.iisc.ernet.in}}

\begin{abstract}
Rotating shear flows, when angular momentum increases and angular velocity decreases as
functions of radiation coordinate, are hydrodynamically stable under linear perturbation. The Keplerian flow
is an example of such systems which appears in astrophysical context. Although decaying eigenmodes
exhibit large transient energy growth of perturbation which could govern
nonlinearity into the system, the feedback of inherent instability to generate turbulence seems questionable. We
show that such systems exhibiting growing pseudo-eigenmodes easily reach an upper bound of growth rate in 
terms of the logarithmic norm of the involved nonnormal operators, thus exhibiting feedback
of inherent instability. This supports the existence of turbulence of hydrodynamic origin in the Keplerian
accretion disc in astrophysics. Hence, this enlightens the mismatch between the linear theory and
experimental/observed data and helps in resolving the outstanding question of origin of turbulence therein.
\end{abstract}


\maketitle

\section{Introduction}
Despite many efforts devoted, the origin of hydrodynamic
turbulence is still poorly understood, because there is a significant mismatch between the
predictions of linear theory and the experimental data (see e.g. \cite{tref,man1}). 
The mismatch between theoretical results and observed data is also found in
astrophysical contexts, where the accretion flow
with Keplerian angular momentum profile is a common subject. 

The hot ionized Keplerian disc flow, which
is linearly stable otherwise, in presence of magnetic coupling is shown to exhibit
magneto-rotational-instability (MRI) \cite{bh91} based on a local analysis. Then the
unstable modes were thought to lead to nonlinearity and then turbulence in the flow. 
However, it was also shown that the criterion drawn from the local dispersion relation is
inadequate and may be misleading \cite{um,mk08}. Then the question arises that
how could a Keplerian accretion disc flow having a very low
molecular viscosity generate turbulence, which could successively govern
diffusive viscosity to support transfer of mass inwards and
angular momentum outwards, in absence of any unstable mode of hydrodynamic origin?

On the other hand, the existence of large transient growth of perturbation in plane 
Couette flow with and without the Coriolis force were demonstrated \cite{tref,man1}. 
It was argued that the large transient growth plausibly could lead to subcritical 
transition to turbulence in laboratory experiments and in astrophysical flows which
in turn could lead to turbulent viscosity \cite{bm08}. 
Note, however, that the shear flows with and without rotation are physically quite different.
The threedimensional growth factors and then plausible turbulent viscosity 
in presence of the Coriolis force were shown to be significant 
in Keplerian flow only if the elliptical vortices present in the background flow \cite{m06,ms10}.
The Coriolis effect, which absorbs the pressure fluctuation, is the main
culprit to kill any growth of energy in the background of linear shear. 
Moreover, very importantly, the feedback of inherent
instability and then sustenance of turbulence based on large transient growth is 
questionable, in lieu of linear instabilities.
Therefore, the problem of hydrodynamic transition to turbulence in rotating shear flow, particularly 
when angular momentum increases and angular velocity decreases as functions of radial coordinate,
e.g. in the astrophysical accretion disc, remains unsolved.
While, Bech \& Anderson
\cite{bech} found turbulence persisting in numerical simulations of
subcritical rotating flows for large enough Reynolds numbers,
in general the problem of hydrodynamic transition 
to turbulence in rotating shear flows, particularly the astrophysical ones, has not been paid enough 
attention in the literature so far.

Decades back Craik \cite{craik} studied the impact of rotation (including the Keplerian one)
and hence the Coriolis effects to the flow
with elliptical streamlines. Later on, Salhi \cite{salhi} examined 
rotation and buoyancy effects on homogeneous shear flows and found similarities between rotation
and stratification effects. In fact, much earlier Bradshaw \cite{brad} discussed the
formal algebraic analogy between the parameters describing streamline curvature and buoyancy
in a turbulent flow. Very recently, the stability problem of unbounded rotating shear flows has been
again studied in presence of uniform vertical density stratification \cite{sc10} which has 
important astrophysical applications e.g. in geometrically thin radially
stratified accretion discs. Note that indeed Johnson and Gammie \cite{jg05} investigated 
nonaxisymmetric radially stratified linearized accretion discs
in a shearing sheet approximation neglecting vertical structure. Then they studied
the time evolution of a plane wave perturbation co-moving with the shear flow.
The shearing flow with rotation and stratification has many
other applications, including that in geophysics, as described in detail by Salhi \& Cambon \cite{sc10}.
Interestingly, same/similar physics is borrowed to understand rotating shear flows in different communities
with the use of different terminologies. While engineering community compute ``rotational 
Richardson number" introduced by Bradshaw \cite{brad} to understand linear stability of a flow
which is proportional to the square of epicyclic frequency used by the astrophysicists
and Rayleigh number by the fluid dynamicists in general for the same purpose. While positive Richardson number 
and Rayleigh number correspond to linear stability in engineering and fluid physics respectively, $q<2$ (real value of 
epicyclic frequency) is the equivalent condition in astrophysics with angular velocity $\Omega\sim r^{-q}$
(will be discussed in detail in subsequent sections).



We show that rotating flows exhibiting pseudo-eigenmodes, 
in absence of any unstable eigenmode itself, quickly 
reach a growth rate of the order of the logarithmic
norm under a very small perturbation. 
This plausibly triggers turbulence. Note that the
logarithmic norm, as discussed later, is an upper bound of such growth rates.
This establishes that the intrinsic perturbation in the system is enough to
sustain the turbulence which is a manifestation of the transient local instability. 
Thus, the turbulence of rotating shear flows and the viscosity of hydrodynamic origin in accretion discs
is shown to be an inherent property of these systems. 

In the next section we describe hydrodynamic equations in shear flows, pseudo-eigenmodes, 
logarithmic norm and corresponding
transient instability in presence of the Coriolis effects. Subsequently, in \S3 we discuss solutions. 
Finally, in \S4 we summarize with conclusions. 

\section{Equations describing shear flow in presence of the Coriolis effect and the corresponding pseudo-eigenmode
and logarithmic norm}

\begin{figure*}
\begin{center}
\includegraphics[scale=0.7]{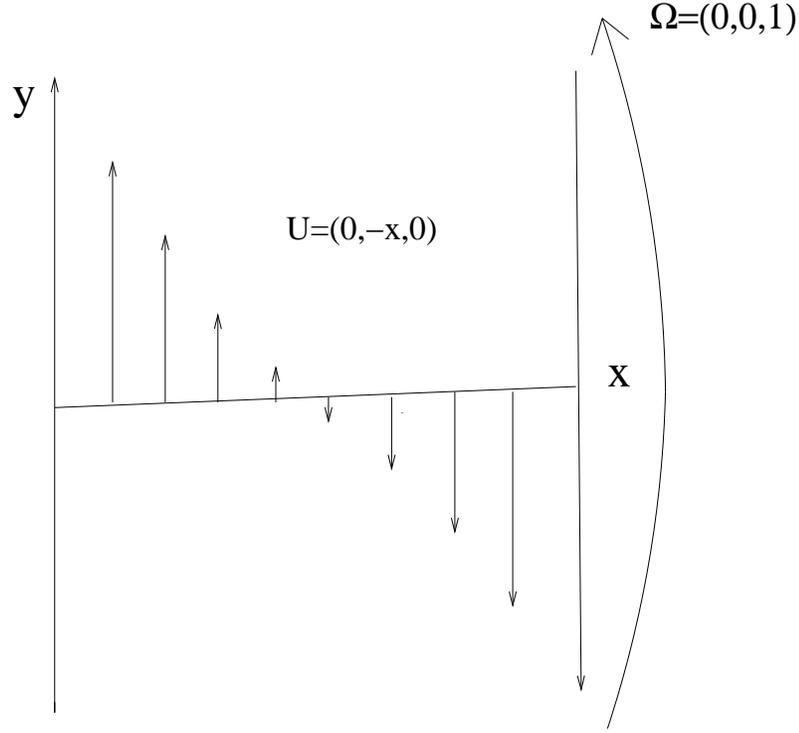}
\end{center}
\caption{Background unperturbed flow in the local comoving box. 
Size of arrows indicates the magnitude of the respective velocities.
} 
\label{shear}
\end{figure*}

Let us focus on a local analysis considering a small patch of rotating shear flow 
whose unperturbed velocity
profile $\vec{U}$ corresponds to the background shear and angular velocity $\Omega\sim r^{-q}$, 
as described earlier \cite{man1}, shown schematically in Fig. \ref{shear}. 
Following earlier work \cite{man1} all the variables are written in the dimensionless form.
The figure describes our choice of local Cartesian coordinates: $x$ is
along the radial direction, $y$ is along the azimuthal or streamwise direction, and $z$ is along 
the vertical direction. The unperturbed flow has a velocity in the $y$-direction and a gradient of velocity
i.e. shear along the $x$-direction. The Coriolis force arised due to rotation of the patch is described by an 
angular velocity vector $\Omega$ along the $z$-direction.

For convenience, we write down the corresponding equations for evolution of perturbation in terms of
the set of velocity and corresponding vorticity $(u,\zeta)$ and hence 
the Orr-Sommerfeld and Squire equations 
in the presence of Coriolis acceleration are given by
\begin{equation}
\left(\frac{\partial}{\partial t}+U\frac{\partial}{\partial y}\right)\nabla^2 u
-\frac{\partial^2U}{\partial x^2} \frac{\partial u}{\partial y}
+\frac{2}{q}\frac{\partial \zeta}{\partial z}
=\frac{1}{R_e}\nabla^4 u
\label{v}
\end{equation}
\begin{equation}
\left(\frac{\partial}{\partial t}+U\frac{\partial}{\partial y}\right)\zeta
-\frac{\partial U}{\partial x} \frac{\partial u}{\partial z}
-\frac{2}{q}\frac{\partial u}{\partial z}=\frac{1}{R_e}\nabla^2 \zeta,
\label{zeta}
\end{equation}
where $u$ is the $x$-component of velocity and $\zeta$ is the corresponding component of vorticity, 
$q=3/2$ for a Keplerian flow and $q=2$ for a constant angular momentum flow, and $R_e$ is the
Reynolds number. Note that in presence of linear background shear, what is of present interest,  
$\vec{U}=(0,-x,0)$ and hence the term proportional to $\frac{\partial^2 U}{\partial x^2}$ vanishes. 
For detailed description, see Mukhopadhyay et al. \cite{man1}. We further reduce the set
of equations in the form of an eigenvalue equation
\begin{eqnarray}
\frac{\partial Q}{\partial t}=-i{\cal L} Q,
\label{sol6}
\end{eqnarray}
where 
\begin{eqnarray}
\nonumber
Q=\left(\begin{array}{cr}\hat{u}\\ 
\hat{\zeta}\end{array}\right),\\\nonumber u(x,y,z,t)=\hat{u}(x,t) \exp[i(k_yy+k_zz)],\\
\zeta(x,y,z,t)=\hat{\zeta}(x,t) \exp[i(k_yy+k_zz)],
\label{sol5}
\end{eqnarray}
and ${\cal L}={\cal L}(k_y,k_z,R_e)$, is a differential operator.
Note that because of translation invariance of the unperturbed flow in the $y$
and $z$ directions, we decompose the perturbations in terms of Fourier modes.



The solution of Eqn. (\ref{sol6}), describing the evolution of perturbation, can be written 
in terms of an eigenfunction expansion
\begin{eqnarray}
\nonumber
&&Q(x,t)=\sum_{j=1}^{\infty} C_j \exp(-i\sigma_j t) {\tilde Q}_j(x),\\
{\rm and}\,\,\,\,\,\,&&{\tilde Q}_j(x)=\left(\begin{array}{cr}\tilde{u}_j(x)\\
\tilde{\zeta}_j(x)\end{array}\right),
\label{sol55}
\end{eqnarray}
where the complex eigenvalue of the problem
$\sigma_j=\sigma_{Rj}+i\sigma_{Ij}$. Therefore, for the $j$th
mode, Eqn. (\ref{sol6}) reduces to
\begin{eqnarray}
{\cal L} {\tilde Q}_j=\sigma_j {\tilde Q}_j.
\label{sol66}
\end{eqnarray}
To obtain the set of eigenvalues and eigenvectors, we convert the
differential operator ${\cal L}$ into an $N\times N$ matrix in a
finite-difference representation and then compute the eigenvalues
and eigenvectors of the matrix. The physical parameters involved
in the problem (mainly $R_e$ and the components of perturbation wave-vector $k_y,k_z$ described below) 
determine the required order $N$ of the matrix
for adequate accuracy. 

It is well known (see e.g. \cite{man1}) that all the eigenmodes are stable, i.e. $\sigma_{Ij}<0$ for all $j$, when $q<2$. 
However, the corresponding pseudo-eigenmodes may be unstable. Following previous work 
discussing nonrotating shear flows \cite{tref},
let us try to understand it. If the linearized fluid system is driven by a signal of the form
$S(x,t)=e^{-i\sigma t}s(x)$, then Eqn. (\ref{sol6}) becomes
\begin{eqnarray}
\frac{d Q}{d t}=-i{\cal L} Q + e^{-i\sigma t}s.
\label{sol7}
\end{eqnarray}
It can be verified that the response will be of the form $Q(x,t)=ie^{-i\sigma t} \varphi(x)$
where $\varphi=(\sigma I - {\cal L})^{-1}s$ and $I$ is the identity matrix.
The operator $(\sigma I - {\cal L})^{-1}$
transforms the input $s$ to the linearized flow at frequency $\sigma$ into the
corresponding output $\varphi$. The maximum amplification rate that occurs in the process can be
written as
\begin{eqnarray}
||(\sigma I - {\cal L})^{-1}||\equiv {\rm maximum}\left(\frac{||\varphi||}{||s||}\right),
\label{amp}
\end{eqnarray}
where $\|. \|$ denotes a norm of the respective quantities, in the present 
work $2-$norm \cite{man1}.

Now an eigenvalue of ${\cal L}$ is $\sigma$ if
\begin{eqnarray}
{\cal L} \varphi =\sigma I \varphi,
\label{eigen}
\end{eqnarray}
where $\varphi$ is the corresponding eigenvector. On the other hand, using Eqns. 
(\ref{amp}) and (\ref{eigen}) $\sigma$ 
can be identified as an eigenvalue of
${\cal L}$ if the perturbation with frequency $\sigma$ brings an unbounded
amplification so that
\begin{eqnarray}
A(k_y,k_z,R)\equiv ||(\sigma I - {\cal L})^{-1}||=\infty.
\label{eigen2}
\end{eqnarray}

Generalizing the above definition of an eigenvalue, we can define $\sigma$ as
an $\epsilon$-pseudo-eigenvalue \cite{trefbk} of ${\cal L}$ for any $\epsilon\ge 0$, if
\begin{eqnarray}
A(k_y,k_z,R)\equiv ||(\sigma I - {\cal L})^{-1}||=\epsilon^{-1},
\label{epeigen}
\end{eqnarray}
when $\varphi$ is the corresponding $\epsilon$-pseudo-eigenvector/mode such that
\begin{eqnarray}
||({\cal L}-\sigma I) \varphi ||=||\epsilon \varphi||.
\label{epeigen2}
\end{eqnarray}
Naturally as $\epsilon\rightarrow 0$, $\varphi$ and $\sigma$ tend to be the pure eigenmode
and eigenvalue respectively. The set of $\epsilon$-pseudo-eigenvalues of an operator
is called the $\epsilon$-pseudo-eigenspectrum.

Therefore, we can classify the amplification in three regimes. First, if 
$R_e$ is equal or greater than a certain critical value $R_{ec}$ 
($R_e\ge R_{ec}$), and
${\cal L}$ has at least one {\it eigenvalue} in the upper half-plane of the Argand diagram,
then $\epsilon=0$ and $A=\infty$;
the flow in this regime is linearly unstable (i.e. the system has at least one
exponential growing mode). Second, if $R_e$ is smaller than a certain value, $R_{eg}$ 
($R_e<R_{eg}$), and all the {\it eigenvalues} lie in the lower half-plane of the Argand diagram,
then $\epsilon=1$ and $A=1$; this implies no amplification (i.e. there is no exponential
growing mode). Finally, if $R_{eg} \le R_e < R_{ec}$
and ${\cal L}$ has at least one pseudo-eigenvalue in the upper half-plane
(although all the eigenvalues still lie in the lower half-plane), then
$1 > \epsilon>0$ and $1<A<\infty$; this is the regime of linearly stable flow with
the existence of transient growth (i.e. the system has at least one growing pseudo-eigenmode).

\subsection{Logarithmic norm and minimum perturbation}
Now a very important related quantity is 
the logarithmic norm which is the upper bound of the possible transient growth rate.
The logarithmic norm, $\mu_p$, of a square matrix operator ${A}$ using a matrix $p$-norm is defined \cite{hairer} 
 as
\begin{eqnarray} 
\mu_p := \lim_{h\to 0, h>0, h \in {\mathbb R}} \frac{\|I + h{A}\|_p -1}{h} \label{mudef} 
\end{eqnarray}
for a natural number $p=1,2, \cdots, \infty$, where $h$ is the time interval, ${\mathbb R}$ is the set of real numbers.  
For $p=2$, it is readily shown that 
\begin{eqnarray} \mu_2({A}) = \lambda_{\max}\left(\frac{{A} +{A}^{\dagger}}{2}\right) \ge \max\left(Re[\lambda ({A})]\right), \label{mu2} 
\end{eqnarray} where $\lambda$ denotes an eigenvalue of the symmetric operand matrix and 
$Re$ indicates real part.  The 
growth of the solution $Q(x,t)$ of
Eqn. (\ref{sol7}) over a non-zero time interval $h$ can be upper bounded \cite{hairer} as follows:
\begin{eqnarray} 
\nonumber
||Q(x,t+h)||_p &\leq&\| I + (- i {\mathcal L}) h \|_p  \|Q(x,t)\|_p \\
&+& h |e^{-i\sigma t}| \|s\|_p + {\mathcal O}(h^2)
\end{eqnarray}
and the limiting maximum growth rate then can be written as 
\begin{eqnarray}
\nonumber
||Q(x,t+h)||_p &\le& \lim_{h\to 0, h>0} \bigg( \frac{ \| I + (- i {\mathcal L}) h \|_p -1}{h}  \|Q(x,t)\|_p \\
&+& |e^{-i\sigma t}| \|s\|_p + {\mathcal O}(h) \bigg) \nonumber \\
&\le&\mu_p(- i {\mathcal L} )  \|Q(x,t)\|_p + |e^{-i \sigma t}|\, \|s\|_p
\end{eqnarray}
from which one can see that the logarithmic norm of $- i {\mathcal L} $ determines the upper bound of the transient growth of the system in Eqn. (\ref{sol7}). Hence, for a suitably small perturbation 
if we obtain a pseudo-eigenvalue which is close to the logarithmic norm and is in the
upper half plane of the Argand diagram, then the system will exhibit sustainable transient instability.
The non-negative real parts in the pseudo-eigenspectrum of  $- i {\mathcal L} $, i.e. the values in the upper half plane of the Argand diagram for the pseudo-eigenspectrum of $ {\mathcal L} $,  drive the transient growth. The pseudo-eigenvalue of  $-i{\mathcal L}$ can be defined, equivalently to Eqn. (\ref{epeigen}), as \cite{trefbk}
\begin{eqnarray}
\lambda_{\epsilon} (-i {\mathcal L}) :=  \lambda(-i {\mathcal L}  + E),~ \|E\|=\epsilon > 0, \epsilon \in {\mathbb R}.
\end{eqnarray}
Using the triangular inequality property of the logarithmic norm, we can write
\begin{eqnarray}
\max\left[Re \big( \lambda_{\epsilon} (-i {\mathcal L}) \big)\right] \le
\mu_2(-i {\mathcal L} + E) \le \mu_2(-i {\mathcal L})  + \epsilon.
\label{lognormeps}
\end{eqnarray}
The above inequality leads to the computation of a minimum $\epsilon$ (namely $\epsilon_{\min}$)
for which the equality holds. If $\epsilon_{\min}$
is small (compared to the matrix $2$-norm of the operator, i.e., $\|-i{\mathcal L}\|_2$ ) such that $\max\left[Re \big( \lambda_{\epsilon} 
(-i {\mathcal L}) \big)\right] > 0$, i.e., there are pseudo-eigenvalues in the upper half plane 
of the Argand diagram, then we can conclude that the rotational shear flow system with the 
Coriolis effect exhibits transient growth sustained by small intrinsic perturbations. 

\section{Optimal perturbations for growing pseudo-eigenmodes and logarithmic norms}
\begin{figure*}
\begin{center}
\includegraphics[scale=0.23]{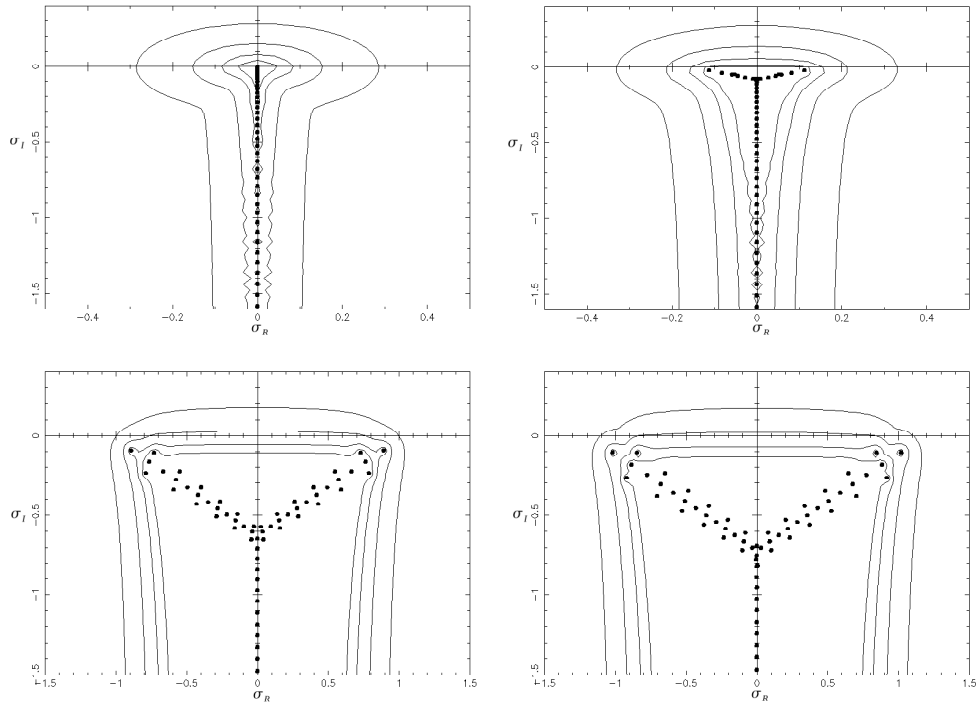}
\end{center}
\caption{Pseudo-eigenspectra of the flow under optimum perturbation
when $R_e=2000$ for $q=2$ (upper-left), $q=1.99$ (upper-right), $q=1.7$ (lower-left), 
$q=1.5$ (lower-right).
Various curves in each figure sequentially from outside to inside respectively
indicate results with $\epsilon=10^{-1}, 10^{-1.5}, 10^{-2}, 10^{-2.5}$.
} 
\label{fig1}
\end{figure*}

We solve the eigenvalue equation (\ref{sol66}) with no-slip boundary conditions
\begin{equation}
u=\frac{\partial u}{\partial x}=\zeta=0,\,\,\,{\rm at}\,\,x=\pm 1.
\end{equation}
In Table 1 we enlist the optimum sets of parameters governing the perturbation that induce maximum growth for various values of $q$. We compute the maximum transient growth $G_{max}$ induced by the
perturbation and the pseudo-eigenvalues determining the 
lower bound of the transient growth of energy: $\mathrm{max}(\left(Im{\left(\sigma\right)}\right)/\epsilon)^2\leq G_{\max}$, where $Im$ indicates imaginary part. Interestingly,
while for $q=2$ the optimum perturbation giving rise to the maximum 
growth is purely vertical, for $q=1.5$ it is 
purely twodimensional. For $1.5< q< 2$ it is threedimensional with $k_y$ and
$k_z$ both nonzero. Figure \ref{fig1} shows the optimum pseudo-eigenspectra for various
values of $q$. Naturally for $q=2$ the spectrum is marginally
unstable and the pseudo-eigenspectra for all values of $\epsilon$ extend
to the upper half plane in the Argand diagram. However, as $q$ decreases, particularly from 
$1.7$ to $1.5$, the pseudo-eigenvalues
for particular values of $\epsilon$, e.g. $10^{-2}$ and $10^{-2.5}$, decrease,
which in turn decrease the rate of maximum growth as shown in Table 1. 

\begin{figure*}
\begin{center}
\includegraphics[scale=0.23]{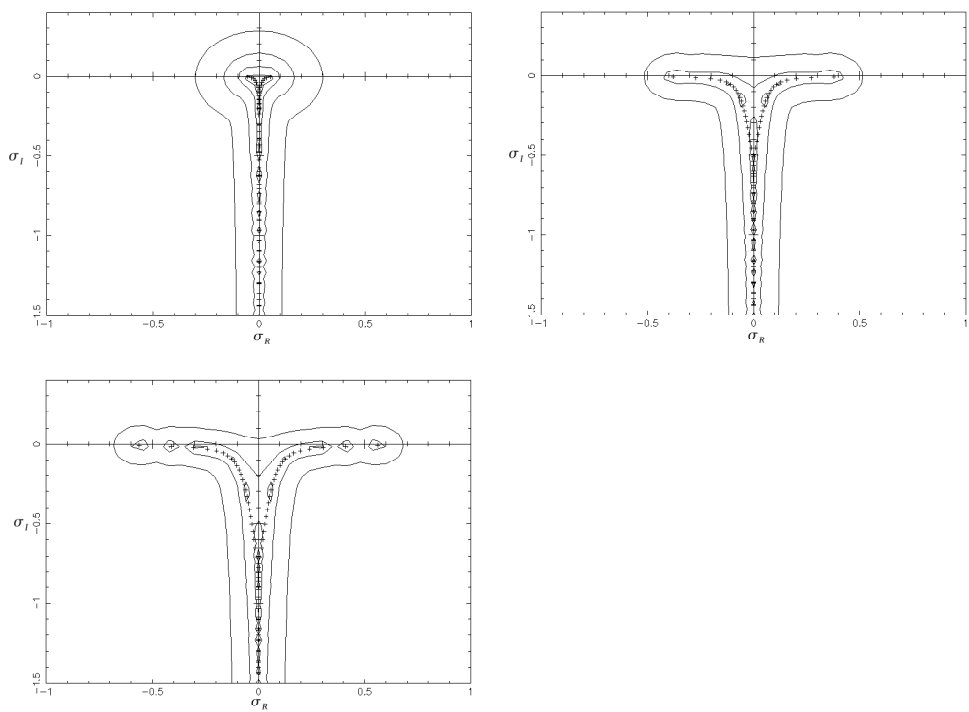}
\end{center}
\caption{Pseudo-eigenspectra of the flow under maximal pure vertical perturbation
when $R_e=2000$ for $q=1.99$ (upper-left), $q=1.7$ (upper-right), $q=1.5$ (lower-left).
Various curves in each figure sequentially from outside to inside respectively
indicate results with $\epsilon=10^{-1}, 10^{-1.5}, 10^{-2}, 10^{-2.5}$.
} \label{fig2}
\end{figure*}

\begin{figure*}
\begin{center}
\includegraphics[scale=0.23]{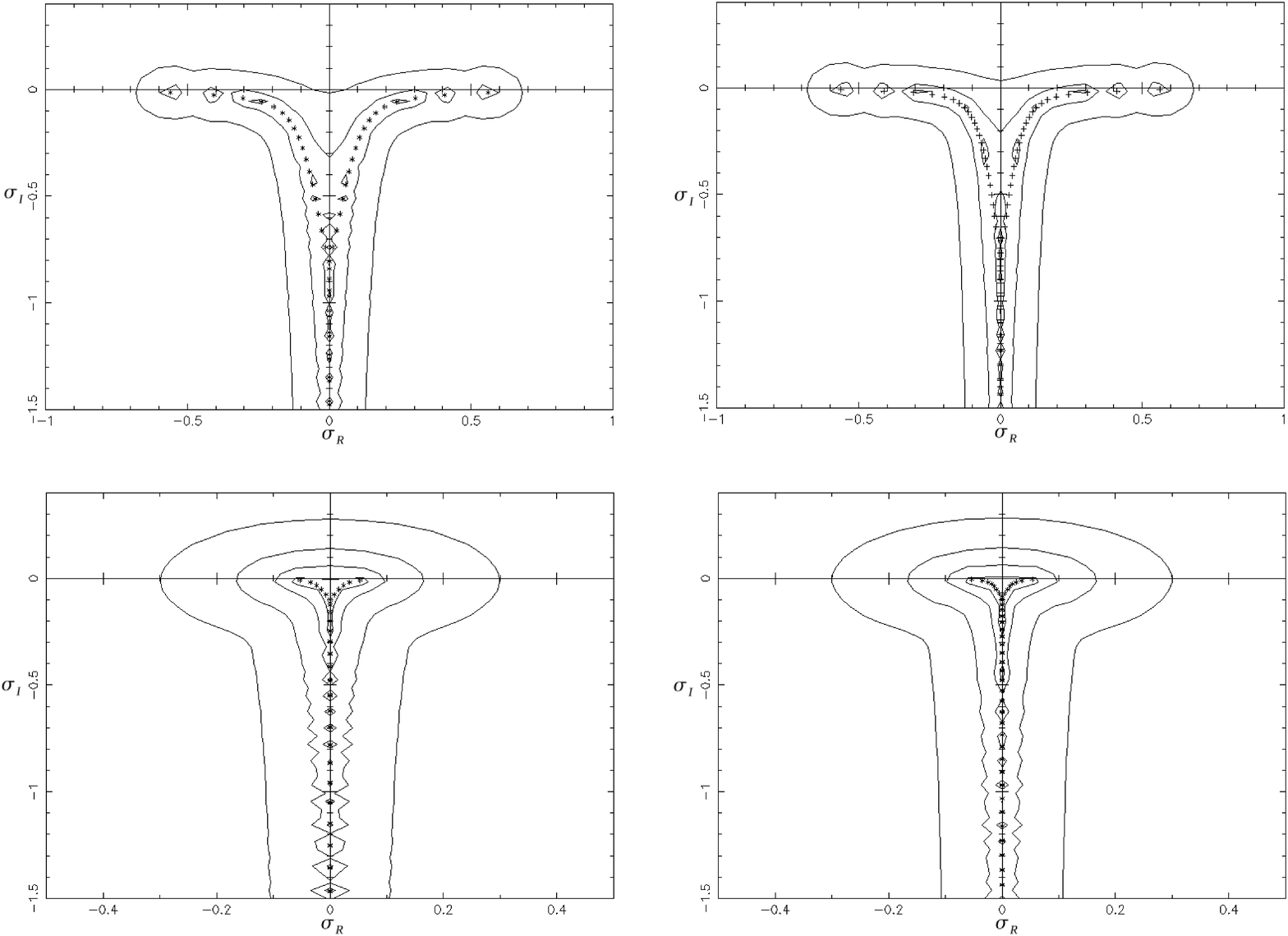}
\end{center}
\caption{Comparison of Pseudo-eigenspectra of the flows under maximal pure vertical perturbation 
between $R_e=1000$ (upper-left), $R_e=2000$ (upper-right), when $q=1.5$; and
$R_e=1000$ (lower-left), $R_e=2000$ (lower-right), when $q=1.99$
Various curves in each figure sequentially from outside to inside respectively
indicate results with $\epsilon=10^{-1}, 10^{-1.5}, 10^{-2}, 10^{-2.5}$.
} \label{fig4}
\end{figure*}

As the maximum growth for $q=2$ is due to the pure vertical perturbation,
we show in Fig. \ref{fig2} the pseudo-eigenspectra due to the pure vertical perturbation,
giving rise to the maximum possible growth (maximal pure vertical perturbation),
for $q<2$ as well. Corresponding parameters are given in Table 2. Interestingly,
the spectra and thus growth for a particular $q$ appear quite different between
that for the optimum and vertical perturbation (particularly for smaller values of $q$). 
This is due to the Coriolis effect generating epicyclic oscillations appeared
in the later case, which is responsible for hindering growing pseudo-eigenmodes.
Note that the energy growth due to pure vertical perturbation is bounded
by the dynamics of the system and is independent of $R_e$. In Fig. \ref{fig4}
we compare pseudo-eigenspectra of the flow with $q=1.5$ and $1.99$ for two different values
of $R_e$ and find them very similar, reflecting their independence of $R_e$
under pure vertical perturbation.
  

Table 3 shows the minimum perturbation ($\epsilon_{\min}$) for which the logarithmic norm rate of growth is approached.
It may be noted that $\epsilon_{\min} \ll \|-i {\mathcal L}\|_2$. Also, at $q=2$ the flow is most susceptible to
transient growth since the required perturbation is the least. A decrease in $q$ reduces the 
susceptibility to
transient growth as the magnitude of the required perturbation increases.
However, $\epsilon_{\min}$ is still quite small even for $q=1.5$, assuring 
local instability and sustained turbulence.

\section{Summary and Conclusions}
While the existence of unstable pseudo-eigenmodes favors instability 
and then plausible turbulence under finite amplitude perturbation,
the question remained whether the local instability is sustained or not.
The present analysis argues that the rotating shear flows with the Coriolis
force not only exhibit unstable pseudo-eigenmodes under suitable small perturbations,
but also the growth rate in the pseudo-eigenmodes approaches 
a strictly positive logarithmic norm and thus assures feedback of
instability as an intrinsic property of the flows. 
Therefore, a very small amplitude of perturbation is able to trigger
instability and then plausible turbulence in the system.
In lieu of linear instabilities (of hydrodynamic and hydromagnetic origin) unstable pseudo-eigenmodes
repopulating the growing disturbance assure {\it sustained} turbulence.
This argues, for the first time, in favor of {\it sustained} 
hydrodynamic instability and then turbulence 
in a rotating Couette like flow, when angular momentum increases and angular
velocity decreases with radius. This is practically the flow of many astrophysical
systems, e.g. accretion disc around compact objects,
some active galactic nuclei, 
star-forming systems. This is possible to conceive
because of the non-self adjoint nature of underlying operator, giving rise
to the non-normal eigenmodes and hence unstable pseudo-eigenspectrum and positive logarithmic norms.
Note that molecular viscosity is too small to explain the observed accretion 
efficiencies in the systems mentioned above.
Therefore, any viscosity is expected to arise due to turbulence in the flow.
Hence, in absence of any linear instability in the Keplerian accretion flows,
existence of small perturbations triggering local transient growth rates of 
the order of logarithmic norm, which is strictly positive for the associated
spatial operator matrix, resolve the question of instability and 
corresponding feedback process which replenish the growth to an ultimate turbulence state. 
This is a new alternate picture of turbulence arousal in astrophysical flows
explaining the transport of mass inward removing angular momentum outward,
particularly when transport due to the nonlinearly developed MRI is expected
to be very small \cite{um} in experimental setups with small magnetic Prandtl number.

\begin{table}
\vskip0.2cm
{\centerline{\large Table 1}}
{\centerline{Parameters for optimal/maximal
perturbation with $R_e=2000$}}
\begin{center}
{
\vbox{
\begin{tabular}{ccccclllllllllllllll}
\hline
\hline
$q$ & $k_y$ & $k_z$ &  $G_{max}$ & $\mathrm{max}\left(\left[Im(\sigma)\right] /\epsilon\right)^2$ &
$\mathrm{max}\left[Im(\sigma)\right]$   \\
\hline
\hline
$1.5$  & $1.2$ & $0$ & $21.67$   & $2.81$ & $0.165$ \\
\hline
$1.7$ & $1.05$ &$0.6$ & $23$   & $2.95$ & $0.159$ \\
\hline
$1.99$  & $0.14$ &$1.64$ & $174.43$   & $38.11$ & $4.285\times 10^{-2}$ \\
\hline
 $2$  & $0$ &$1.66$ & $4661$   & $ 2199.96$ & $3.56\times 10^{-3}$ \\
\hline
\hline
\end{tabular}
}}
\end{center}


\vskip0.2cm {\centerline{\large Table 2}} {\centerline{
Parameters for maximal pure vertical
perturbation with $R_e=2000$}}
\begin{center}
{
\vbox{
\begin{tabular}{ccccllllllllllllll}
\hline
\hline
$q$ & $k_z$ &  $G_{max}$ & $\mathrm{max}\left(\left[Im(\sigma)\right] /\epsilon\right)^2$  & 
$\mathrm{max}\left[Im(\sigma)\right]$\\
\hline
\hline
$1.5$  & $2.5$ & $3.84$   & $1.54$ & $0.945$ \\
\hline
$1.7$ & $2.4$ & $6.31$   & $2.15$  & $0.516$\\
\hline
$1.99$  & $1.9$ & $153.4$   & $42.74$  & $5.419\times 10^{-2}$\\
\hline
 $2$  & $1.66$ & $4661$   & $2199.96$ &  $3.56\times 10^{-3}$\\
\hline
\hline
\end{tabular}
}}
\end{center}

{\centerline{\large Table 3}}
{\centerline{Minimum perturbation for logarithmic norm growth rate 
at $R_e=2000$ }}
\begin{center}
{ 
\vbox{
\begin{tabular}{ccccclllllllllllllll}
\hline
\hline
$q$ & $k_y$ & $k_z$ &  $\mu_2(-i{\mathcal L})$ & $\epsilon_{\min}$ & $\|-i {\mathcal L}\|_2$ \\
\hline
\hline
$1.5$  & $1.2$ & $0$ & $1.1830$ & $ 4.05\times 10^{-2}$ & $5.1063$\\
\hline
$1.7$ & $1.05$ &$0.6$ & $1.0430$ & $1.05\times 10^{-2}$ & $ 5.0079$ \\
\hline
$1.99$  & $0.14$ &$1.64$ &  $0.3851$  &$1.98\times10^{-3} $ & $5.0089$ \\
\hline
 $2$  & $0$ &$1.66$ & $0.3632$   & $1.86\times 10^{-5}$ & $5.0087$  \\
\hline
\hline
\end{tabular}
}}
\end{center}

\end{table}


\section*{Acknowledgments}
The authors would like to thank the referee for his/her comments which helped 
to improve the presentation of the paper.
Two of the authors (BM, RM) were supported in part by DST grant SR/S2HEP12/2007 funded by
government of India. Another author (SR) was partially supported by
the IISc Mathematics Initiative (IMI) programme.

\section*{References}

\end{document}